\documentclass[compsoc, conference, a4paper, 10pt, times]{IEEEtran}
\usepackage{cite}
\usepackage{amsmath,amssymb,amsfonts}
\usepackage{algorithmic}
\usepackage{graphicx}
\usepackage{textcomp}
\usepackage{xcolor}
\usepackage{booktabs}
\usepackage{url}

\usepackage{breakurl}
\usepackage[hidelinks, breaklinks]{hyperref}
\usepackage[inline]{enumitem}
\usepackage{lipsum}
\usepackage{listings}
\usepackage{svg}

\begin{document}

\title{Your Vulnerability Disclosure Is Important To Us: An Analysis of Coordinated Vulnerability Disclosure Responses Using a Real Security Issue}

% Submissions should be anonymized. See the CFP for details on how to anonymize your paper, including any references to your own work.
%\author{\em Anonymous Authors}
%TODO ANON

% The author information is skipped here, but can be used to include author information in the publication.
%\iffalse
\author{\IEEEauthorblockN{Koen van Hove}
\IEEEauthorblockA{\textit{NLnet Labs \& University of Twente} \\
Amsterdam, the Netherlands \\
koen@nlnetlabs.nl}
\and
\IEEEauthorblockN{Jeroen van der Ham-de Vos}
\IEEEauthorblockA{\textit{University of Twente} \\
Enschede, the Netherlands \\
j.vanderham@utwente.nl}
\and
\IEEEauthorblockN{Roland van Rijswijk-Deij}
\IEEEauthorblockA{\textit{University of Twente} \\
Enschede, the Netherlands \\
r.m.vanrijswijk@utwente.nl}
%% IEEE format can accommodate up to six authors this way
}
%\fi

\maketitle

\begin{abstract}
	It is a public secret that doing email securely is fraught with challenges. We found a vulnerability present at many email providers, allowing us to spoof email on behalf of many organisations. As email vulnerabilities are ten a penny, instead of focusing on yet another email vulnerability we ask a different question: how do organisations react to the disclosure of such a security issue in the wild? We specifically focus on organisations from the public and critical infrastructure sector who are required to respond to such notifications by law. We find that many organisations are difficult to reach when it concerns security issues, even if they have a security contact point. Additionally, our findings show that having policy in place improves the response and resolution rate, but that even with a policy in place, half of our reports remain unanswered and unsolved after 90~days. Based on these findings we provide recommendations to organisations and bodies such as ENISA to improve future coordinated vulnerability disclosure processes.
\end{abstract}

% Depending on how vigilant their paper processor is, IEEE may ask for these in your final paper, but we've heard about these amazing inventions called search engines that are able to index every word in your paper, so no need to include them in your submission unless you really want to.

%\begin{IEEEkeywords}
%component, formatting, style, styling, insert
%\end{IEEEkeywords}

\section{Introduction}
Email security is notoriously difficult. Security measures come in the form of a patchwork of protocols that are sometimes difficult to interpret and implement. It is therefore no surprise that the community regularly identifies serious email vulnerabilities. In this paper, we study `yet another' email vulnerability. Rather than focusing on `yet another' vulnerability, however, we focus on a different aspect of vulnerabilities: the disclosure process, and how organisations deal with disclosures.

The issue we uncovered impacts the authenticity of email, allowing us to spoof email from many organisations. The issue impacts governments and critical infrastructure in the Netherlands, Belgium, and other countries throughout Europe. Traditionally, the sender of an email has been as trustworthy as the sender of a letter. Everyone can write a physical letter supposedly coming from `10 Downing Street', although in most cases that will be incorrect. With the advent of spam and phishing, methods to prevent this for electronic mail were created to solve that, from which a number of standards to verify the sender of email emerged: the Sender Policy Framework (SPF) \cite{rfc7208}, DomainKeys Identified Mail (DKIM) \cite{rfc6376, rfc8301, rfc8463} and Domain-based Message Authentication, Reporting \& Conformance (DMARC) \cite{rfc7489}. These technologies have not been without controversy. SPF in particular was claimed to ``break the internet'', particularly verbatim email forwarding \cite{weng_2004}, whereas DMARC caused issues for mailing lists \cite{rfc7960}. Nevertheless, adoption and support for these standards is rising.

In 2019, these three technologies became mandatory to implement for Dutch governmental bodies and critical infrastructure, based on a decision by Forum Standaardisatie (The Standardisation Forum of the Netherlands) to adopt them as `must implement' standards \cite{forumstandaardisatie_spf, forumstandaardisatie_dkim, forumstandaardisatie_dmarc}. Adoption is significant, with for example in May~2022 89\% of Dutch municipalities fully using all standards correctly~\cite{forumstandaardisatie_2022}. Certain suppliers also voluntarily signed a manifest promising to adhere to the guidelines~\cite{forumstandaardisatie_leveranciersmanifest}. In addition to this, secure email is also a `technical agreement' for health care purposes in the Netherlands as defined in NTA~7516 \cite{nta7516}.

We use this opportunity to study how organisations who are mandated to securely implement these standards, and are also affected by the vulnerability we uncovered, react to disclosure. In a survey from the National Telecommunications and Information Administration (NTIA), over half of security researchers have indicated experiencing frustration during disclosures \cite{ntia_2016}. The European Commission seems to share this, and to counteract that has introduced the Cybersecurity Act \cite{eu-881-2019} in 2019, which gives the European Network and Information Security Agency (ENISA) the mandate to create the EU Cybersecurity Certification (EUCC) \cite{enisa_2021}, of which the latest version stems from 2021. This certification is likely to become mandatory-to-implement for certain organisations in future legislation, such as the upcoming Network and Information Systems guideline (NIS2) \cite{eu-333-2022}. %, or as part of national law.

The question we ask ourselves is two-fold: \textit{How do organisations respond to our coordinated vulnerability disclosure (CVD) notification and how does their handling of the disclosure align with their policy?}

Our main contributions are that we
\begin{enumerate*}
	\item describe a novel email vulnerability and analyse its prevalence among organisations in the public sector in 10~countries;
	\item perform a large scale CVD to public sector organisations in the Netherlands and Belgium in order to analyse and quantify how these notifications are treated;
	\item examine the policy of these organisations, and how that correlates with the responses we see in practice.
\end{enumerate*}

\vspace{0.5em}
\noindent
\textbf{Outline --} 
The remainder of this paper is structured as follows: in Section~\ref{sec:background} we provide background information on email standards and discuss related work. Section~\ref{sec:exploit} describes the discovered weakness in common configurations of email security. In Section~\ref{sec:measurement-analysis} we analyse the prevalence of this vulnerability. Then, in Section~\ref{sec:disclosure} we outline the process we used to disclose the vulnerability to the affected organisations, and present an overview of the responses and response times. In Section~\ref{sec:policy} we compare our observations in the wild to what should happen according to the organisation's own policies. In Section~\ref{sec:discussion} we discuss our findings and the wider implications of these results. Section~\ref{sec:recommendations} provides recommendations for future coordinated vulnerability disclosure policy. We discuss the ethical considerations of our work in Section \ref{sec:ethics}. Finally, Section~\ref{sec:conclusion} draws conclusions.

\section{Background \& Related Work}\label{sec:background}
 
Our work analyses what happens when one discloses a vulnerability, in this case one concerning email. This section is therefore split up in two parts: a part about vulnerability disclosure, and a part about email security. We discuss related work in the context of both parts.

\subsection{Vulnerability Disclosure}
{\it Vulnerability disclosure} is the act of providing information about a functional behaviour of a product or service that violates an implicit or explicit security policy to a party that was not believed to be previously aware \cite{iso29147:2018}. {\it Coordinated} vulnerability disclosure (CVD) is a vulnerability disclosure process that includes coordination, that is, it includes identifying and engaging stakeholders, mediating, communicating, and other planning in support of the disclosure process.

Interestingly, neither of the two best known information security standards, ISO/IEC 27001:2022~\cite{iso27001:2022} and 27002:2022~\cite{iso27002:2022}, requires there to be a process for handling vulnerability notifications from third parties. There is, however, a standard specific for vulnerability disclosure, ISO/IEC 29147:2018 \cite{iso29147:2018}, which is also mentioned and included in adapted form as part of the `Cybersecurity Certification: Candidate EUCC Scheme' published by the European Union Agency for Cybersecurity ENISA \cite{enisa_2021}. This standard states that meaningful acknowledgement of receipt of potential vulnerability reports should occur within 7~calendar days, and mandates providing and publishing a contact mechanism, although the specific mechanism is unspecified. Cybersecurity Certification in its current form is voluntary unless otherwise specified in other EU law or national law \cite{eu-881-2019, eu-333-2022} -- the first assessment of which products will be subject to mandatory certification shall be carried out by 31 December 2023. 
Cybersecurity certification schemes such as the EUCC are also referenced in the NIS2 directive as something that EU member states may require entities to comply with \cite{eu-333-2022} -- the NIS2 itself does not directly mandate specific entities to have this certification. The upcoming Cyber Resilience Act also mandates manufacturers to have a coordinated vulnerability disclosure process, but does not specify a timeline \cite{eu-454-2022} within which disclosures must be handled. The U.S.\ Cybersecurity and Infrastructure Security Agency (CISA) uses a timeline of 45~days for disclosure if a manufacturer does not reply \cite{cisa_2023}.

When it comes to coordinated vulnerability disclosure policy on a national level, both the Netherlands and Belgium are considered countries at the forefront of policy according to an ENISA report \cite{enisa_2023}. When it comes to legality, both countries take a slightly different approach: whereas in the Netherlands the public prosecutor has the ability to choose whether to prosecute or not and has created policy to not do so in the case of coordinated vulnerability disclosure, in Belgium this is actually enshrined in law~\cite{be-2022034749, be-2022042980}. A big caveat of the Belgian approach is that the national Computer Security Incident Response Team (CSIRT), CERT.be, needs to give permission to make the information public. We note that our disclosures in Belgium pre-date this law coming into effect.

The Dutch government has committed itself to implementing the `Baseline Informatiebeveiliging Overheid' (Baseline Information security Government, BIO), which requires organisations within the government on all levels (central and local) to implement and publish a coordinated vulnerability disclosure procedure \cite{bio_2020}. The NCSC-UK recommends using `security.txt'~\cite{rfc9116} as one of the methods \cite{ncscuk_2020}, something now also required in the Netherlands by Forum Standaardisatie \cite{forumstandaardisatie_securitytxt}.

\vspace{0.5em}
\noindent
\textit{Related work -- }
Research has been conducted into how organisations handle vulnerability disclosures currently. 
Bolz et al.~\cite{bolz_2021} look into that aspect for the automotive industry. Donatello et al.~\cite{donatello_2019} look at the patterns of bug bounty programmes such as HackerOne and Google Vulnerability Research.

Other articles look into the effect of vulnerability disclosures, such as Arora et al.~\cite{arora_2006}, Cavusoglu et al.~\cite{cavusoglu_2007}, and B\"{o}hme~\cite{boehme_2006} who look into what the impact is of different modes of vulnerability disclosure. ENISA analysed the different national vulnerability programmes in the European Union \cite{enisa_2023}.

We aim for a more holistic view -- how does an organisation handle a vulnerability disclosure that is caused by a third-party service? Be it due to a misconfiguration or a problem at the third party they are unwilling to fix.

\subsection{Email}
\subsubsection{SMTP}
The Simple Mail Transfer Protocol (SMTP) was standardised in 1981 \cite{rfc788}. Initially, SMTP transport was unencrypted. Messages were altered by different relays and mail transfer agents~\cite{rfc2476}, and authenticity could not be guaranteed. SMTP authentication was added in an extension in 1999~\cite{rfc2554}. 

SMTP only cares about the transport of the message, not about the content of the message. This means that the SMTP recipient and sender~\cite{rfc5321} and the `from' and `to' address in the Internet Message Format (IMF) of the content of the email~\cite{rfc5322} may be different. For example, when plain forwarding a message from someone else, one may leave the IMF the same but change the SMTP recipient to the new recipient.

\subsubsection{SPF}
The Sender Policy Framework (SPF) was initially published as an experimental standard in 2006~\cite{rfc4408}, and formally in 2014~\cite{rfc7208}. It allows a sender to define IP addresses (or mechanisms that can be resolved to IP addresses) that are authorised to send email on behalf of a domain name. This is the same domain name as the SMTP FROM as defined in RFC~5321~\cite{rfc5321}, and not the header From as defined in RFC~5322~\cite{rfc5322}. It is defined as a TXT record in the Domain Name System (DNS). There are four possible results when checking whether the domain name used in the SPF FROM matches with the SPF record, namely `pass', `neutral', `soft fail' and `fail'. 

% An SPF record starts with ``v=spf1'', and contains one or more mechanism, generally ending with an ``all'' mechanism. Examples of mechanisms are `a', which matches the A or AAAA record, or `mx', which matches the MX records. Every mechanism may be preceded with a qualifier, which states how an IP address that is part of that range should be handled. The lack of a qualifier means that it is a pass. In practice the qualifier is only used for the `all' mechanism, where it is used to fail if it does not match one of the previous mechanisms. There is a standardised limit of at most 10 DNS lookups - if exceeded, SPF must be considered a fail. In practice implementations are more lenient. An example SPF record could look like this: ``v=spf1 mx a:example.org ?ip4:1.2.3.4 -all''. Here the MX record of the current domain is included, as well as the A and AAAA records of example.org. 1.2.3.4 will result in a neutral result, and if the IP address does not match any of the other mechanisms, it will be a fail as indicated by the minus before `-all'.

To ease the management of the SPF record, it is also possible to use the `include' mechanism. This mechanism looks up the SPF record at a different domain, which is consequently parsed in the same way. 

\vspace{0.5em}
\noindent
\textit{Related work -- }
%The security of SPF has been looked into from an academic perspective, such as SPFail by 
Bennett et al.~\cite{bennett_2022}, studied a vulnerability in a popular SPF parser and measured the number of vulnerable instances over time. Sipahi et al.~\cite{sipahi_2015} looked into whether the way the SPF record was constructed could be used to determine spam domains. 

Some non-peer-reviewed work can also be found in this field, such as by Salla~\cite{salla_2022}, who describes an attack where the SPF record of several Australian organisations included a publicly available range used by AWS, allowing everyone with a virtual private server in that range to send email on behalf of those organisations. Another example is Bencteux et al.~\cite{bencteux_2022}, who performed a similar experiment in Denmark, where the SPF include of a large webhosting organisation included all their VPSes.

The original standard also looks at how the SPF record could be abused by others to execute a denial-of-service attack or to amplify a DNS attack~\cite{rfc7208}.

\subsubsection{DKIM}
DomainKeys Identified Mail (DKIM) was initially published in 2007~\cite{rfc4871}, and revised in 2009~\cite{rfc5672} and in 2011~\cite{rfc6376}. Unlike SPF, DKIM does not care about IP addresses or SMTP. DKIM signs (parts of) the IMF (i.e.\ the message inside the envelope) with a cryptographic signature, and adds that as a header to the IMF. In order to verify the signature, the recipient can use the domain and selector specified in the DKIM header and retrieve the public key for the signature by making a DNS TXT request to \verb|{selector}._domainkey.{domain}|. 

The resulting TXT record contains both the public key and the algorithm used. Because the headers and body are signed, tampering with those by third-parties can be detected. The domain used to sign the message does not need to correspond to the domain used to send the email (either the SMTP FROM \cite{rfc5321} or header From~\cite{rfc5322}). The guarantee is only as strong as the trust in the signing party. Signing the email with a DKIM key from the sender's domain is however recommended.

\vspace{0.5em}
\noindent
\textit{Related work -- }
There is little work regarding DKIM. The work done motivates what the threats were without DKIM, such as Fenton shows in RFC 4686~\cite{rfc4686}, measurement of its adoption, such as Chuhan et al.~\cite{chuhan_2022}, or solutions for problems such as mailing lists changing the headers, as Higashikado et al.\ do~\cite{higashikado_2008}.

\subsubsection{DMARC}
SPF adds a check for the IP address of the sending SMTP server for the SMTP FROM address, and DKIM allows for signing of the message itself. Both of these combined are not yet sufficient to prevent the spoofing of emails. For example, one could send an email with a different Header from (which is the field generally shown by email clients), and sign it using a different domain. Domain-based Message Authentication, Reporting \& Conformance (DMARC), standardised in 2015~\cite{rfc7489}, tries to do three things: 
\begin{enumerate}
	\item Bridge the gap between SMTP FROM and Header from, or the DKIM domain and Header from, by checking for alignment;
	\item Add reporting to support analysis of the impact of the SPF, DKIM, and DMARC configuration for a specific domain;
	\item Specify policy for the recipient what should be done with email that does not pass the DMARC checks.
\end{enumerate}
The DMARC record can be obtained by making a DNS TXT request for \verb|_dmarc.{domain}|.
This record applies to the domain itself as well as all its subdomains.
DMARC builds upon SPF and DKIM. The way to pass the DMARC check is:
\begin{verbatim}
	(SPF pass AND SPF aligned)
	OR
	(DKIM pass AND DKIM aligned)
\end{verbatim}
By adding alignment, DMARC requires the SMTP FROM to match the Header from in the case of SPF, or the DKIM domain to match the Header from, thereby preventing the spoofing method that was previously possible with only SPF and DKIM.

%Additionally, DMARC includes a reporting standard that allows XML reports to be emailed to a specified email address in the DMARC record. This report contains, among other things, the IP address of the sending SMTP server and the domain used to DKIM sign the email. It also contains which policy the receiving party applied. This allows operators to get better insight into whether their email is delivered and whether their domain is used to send spam.

Whilst an SPF fail or broken DKIM signature are good indications of something not being right, both standards do not specify what a recipient must do with that information. DMARC adds a policy to this, allowing a sender to specify that either nothing should be done, the email should be treated as spam (but still delivered), or the email should be outright rejected.

\vspace{0.5em}
\noindent
\textit{Related work -- }
Most related work regarding DMARC has used DMARC as a means to accomplish something else, such as Kitagawa et al.~\cite{kitagawa_2016} who use it to design a system to notify the receiver of the verification result, and Kanako et al.~\cite{kanako_2020} who use it to do false positive detection.

%\subsubsection{BIMI}
%Brand Indicators for Message Identification (BIMI) is an internet draft currently already supported by for example Google and Yahoo \cite{brand-indicators-for-message-identification-03} to display a brand image next to an email. It is structured a lot like DKIM and DMARC, and also builds upon DKIM and DMARC. The email itself has a header called ``BIMI-Selector'', which might contain a selector much like DKIM (if the selector is not set it defaults to ``default''). Unlike DKIM, specifying another domain is not allowed. The ``BIMI-Location'' header contains a URL to an image. By querying the DNS of the sender's domain using the selector, e.g. `default.\_bimi.example.org', one can obtain a TXT record that verifies that the URL in the BIMI-Location header is allowed to be used for this domain. 

%Part of the BIMI-Location header is a Verified Mark Certificate (VMC), which like WebPKI provides authenticity that this is indeed your brand identity using a set of trusted authorities, by having a certificate with the Verified Mark extension that contains the brand as SVG. This is technically optional. One of the requirements of BIMI is that DMARC is set up using a policy of quarantine or reject.

\subsubsection{Full Stack}
We finally discuss related work that looks at the entire email stack.
Most research into SPF and DMARC looks into its adoption, such as Hu et al.~\cite{hu_2018} and Maroofi et al.~\cite{maroofi_2021}, and whether they are properly configured. Draper et al.~\cite{draper_2022} combine several different measurements into email security protocol adoption. Shen et al.~\cite{shen_2021} look into email spoofing from the recipient side, i.e. what (maliciously) malformed email a client or server will accept. 

The most similar academic work is by Liu et al.~\cite{liu_2023}, who spoof email addresses protected by SPF, DKIM, and DMARC by abusing forwarding vulnerabilities at common email providers. However, in this case the behaviour was clearly unintended and most providers were interested in solving the issue. The most similar non-academic work is by Salvati~\cite{salvati_2023}, who describes the story of reporting a similar issue to what we found to a single transactional email service provider. 

What we aim to do is expand on that by looking at the general sending restrictions at popular web hosting providers for whom email is not their core business and who may have legacy reasons to allow broader sending domains, as well as their impact on spam and phishing email delivery.

\section{Email Vulnerability}\label{sec:exploit}

In this section we elaborate on the issue we found. The core of the issue relates to how shared web hosting parties -- that handle mail on behalf of many different organisations -- configure their outgoing mail servers. What makes them different from the email security vulnerability cases named in section \ref{sec:background} is that email is not their core business -- web hosting is. Shared web hosting parties often share the outgoing email servers between all their customers. This means that from the point of view of SPF, it is the same server and IP address, and thus authorised. The operator of the outgoing mail server thus has to ensure that a customer can only send email on behalf of themselves. 
If the operator does not implement this check, then this means that any customer of that web hoster can send valid email on behalf of other customers.

\begin{figure}
	\begin{lstlisting}[language=bash, basicstyle=\ttfamily, showstringspaces=false]
abc.nl.	1800	IN      TXT     "v=spf1 include:spf.hosting.nl ... -all"
	\end{lstlisting}
	\centering
	\caption{Example SPF record for abc.nl authorising everything included in spf.hosting.nl to send email on behalf of them.}
	\label{fig:dns-spf}
\end{figure}

In figure \ref{fig:dns-spf} we show how the customer's domain \verb|abc.nl| includes \verb|spf.hosting.nl| in its SPF record. This includes everything from \verb|spf.hosting.nl|, which includes all their outgoing mail servers. Other customers, for example \verb|def.nl|, also include \verb|spf.hosting.nl|, and also have access to those same mail servers. If the mail server does not check that the customer sends email on behalf of themselves, rather than from another customer, then the email sent by the other customer is indistinguishable from email sent by the actual holder of the domain name.

\subsection{Methodology}
%For our testing, we adhered to the ethical considerations we describe in Section \ref{sec:ethics}.

We next describe the process through which we test if this vulnerability occurs; we note that we discuss ethical considerations of our work in Section~\ref{sec:ethics}.
The first step in this process is to purchase web hosting and/or email hosting at the same place as the victim does its email hosting. Most web hosting also includes some form of email hosting by default, be that as actual inboxes or merely the sending part from contact forms. Once that has been accomplished, we try the following things in order:

\vspace{0.5em}
\noindent
1) In many shared web hosting environments, a PHP environment is available. We can use this to run code on the web hoster's servers, which may not require the same level of authentication to the outgoing mail servers as a request from outside the web hoster's network does. In those cases, a simple script as shown in figure \ref{fig:php-mail} may be enough. This script simply requests the mail server configured in the server's PHP configuration to deliver an email on behalf of @victim.nl.

\begin{figure}
	\begin{lstlisting}[language=php, basicstyle=\ttfamily, showstringspaces=false]
$to   = $_POST["to"];
$subj = $_POST["subject"];
$msg  = $_POST["message"];
$from = $_POST["name"] . "@victim.nl";
$hdrs = [
  "From" => $from,
  "Reply-To" => $from
];
$add = "-f " . $from;
// mail() is a built-in PHP function
mail($to, $subj, $msg, $hdrs, $add);  
	\end{lstlisting}
	\centering
	\caption{Example PHP code to send email on behalf of @victim.nl. If @victim.nl has the hoster's mail servers in its SPF record, and the hoster does not check whether a customer is allowed to send email on behalf of a certain domain, then running this script on the hoster's web servers will allow email spoofing on behalf of @victim.nl.}
	\label{fig:php-mail}
\end{figure}

\vspace{0.5em}
\noindent
2) Sometimes the mail server used by default by PHP's \textit{mail()} function may reject email from incorrect domains, but elsewhere on the network there is an open relay available to the network. In that case, PHP can be used to directly connect to that open relay over SMTP, for example using a third-party library such as PHPMailer, as shown in figure \ref{fig:php-openrelay}. Here we connect to \verb|relay.example.org|, a relay server provided by the web hoster for use on its servers.

\begin{figure}
	\begin{lstlisting}[language=php, basicstyle=\ttfamily, showstringspaces=false]
$mail = new PHPMailer();  
$mail->isSMTP();
$mail->Host     = "relay.example.org";
$mail->SMTPAuth = false;
$mail->setFrom("spoofed@victim.nl");
...
$mail->send();
	\end{lstlisting}
	\centering
	\caption{Example PHP code to send email on behalf of spoofed@victim.nl using an open relay from a hoster. Setting the recipient, subject, and body has been redacted.}
	\label{fig:php-openrelay}
\end{figure}

\begin{figure}
	\includegraphics[width=1.0\linewidth]{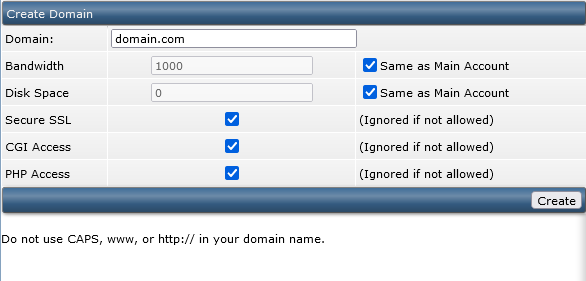}
	\centering
	\caption{Example of the new domain configuration page in DirectAdmin. Screenshot taken from a large Dutch web hoster.}
	\label{fig:directadmin}
\end{figure}

\vspace{0.5em}
\noindent
3) In other cases the domain name used to send email from may be checked (and a domain name not belonging to the account may be rejected), but the adding of new domain names is unrestricted. In figure \ref{fig:directadmin} is a screenshot from a large web hoster in the Netherlands that uses the popular DirectAdmin interface for its customers. One can enter any domain name without checks. Whilst of course the website part does not work as the DNS does not point to the web hoster, it is in many cases still possible to create a mailbox. This mailbox can then be used to send email using the victim's domain.

The precise details differ from hoster to hoster. For example, some web hosters use a separate portal to add new domain names, whereas others use the one built-in to common administration interfaces such as DirectAdmin or cPanel. Additionally, for all these attacks we presume that someone who wishes to send phishing is capable of using a payment method not linked to their identity (e.g., a stolen credit card or a prepaid gift card) to buy the web hosting, or abuse a vulnerable website hosted on one of the platforms.

\subsection{Heuristic}
\begin{figure}
	\includegraphics[width=1.0\linewidth]{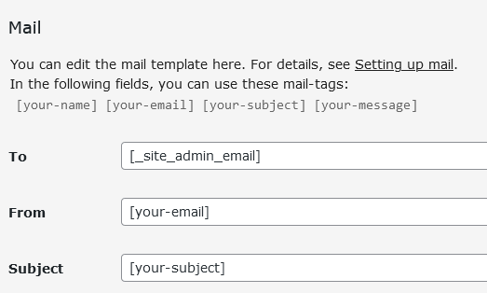}
	\centering
	\caption{Example of the contact form configuration page for a popular contact form plugin for WordPress. The shown configuration does not work if a web hoster checks for the sending email domain.}
	\label{fig:wordpress}
\end{figure}

For a hoster to be vulnerable, it needs to adhere to the following criteria:
\begin{enumerate}
	\item Be able to send email through its own infrastructure -- hosters such as Google Cloud Platform and Microsoft Azure do not allow anyone to send email from their infrastructure (or only with specific uncommon agreements);
	\item Not check what the sending domain is \textit{or};
	\item Not check when adding a domain whether the holder of a domain authorised it;
\end{enumerate}

Some hosters clearly state on their website that they block this behaviour. For example, TransIP, a popular web hoster in the Netherlands, lists the following on their website~\cite{transip_2023} (translated):
\begin{quote}
	To ensure the security and stability of our web hosting platform, it is only possible to send mail from your website if the 'From' address of your contact form contains your domain name. This prevents third parties from sending spam via your website. If you have a PHP script to use a contact form on, say, yourdomain.nl, it is necessary for the sender address to end in @yourdomain.nl. This is usually sufficient to use a contact form.
\end{quote}
Similar texts can be found on, for example, Hostnet.nl, One.com, and Domeneshop.no. The absence of such a text generally provides a strong indication that abuse is possible. Whilst we do not know for certain why web hosters add a text like the one from TransIP to their website, we can speculate. Imagine one has a website with a contact form with the usual name, email, subject, and message fields. It is very tempting to add the respondent's email address as the email's `From' address, as for example shown in Figure~\ref{fig:wordpress}. Filtering on the domain would break these websites, and the owner of the website may decide to blame the hoster (rather than the person who developed their website). We have had confirmation from at least one web hoster that this was indeed the case.

\section{Measurement Analysis}\label{sec:measurement-analysis}
\begin{figure}[]
	\includegraphics[width=1\linewidth]{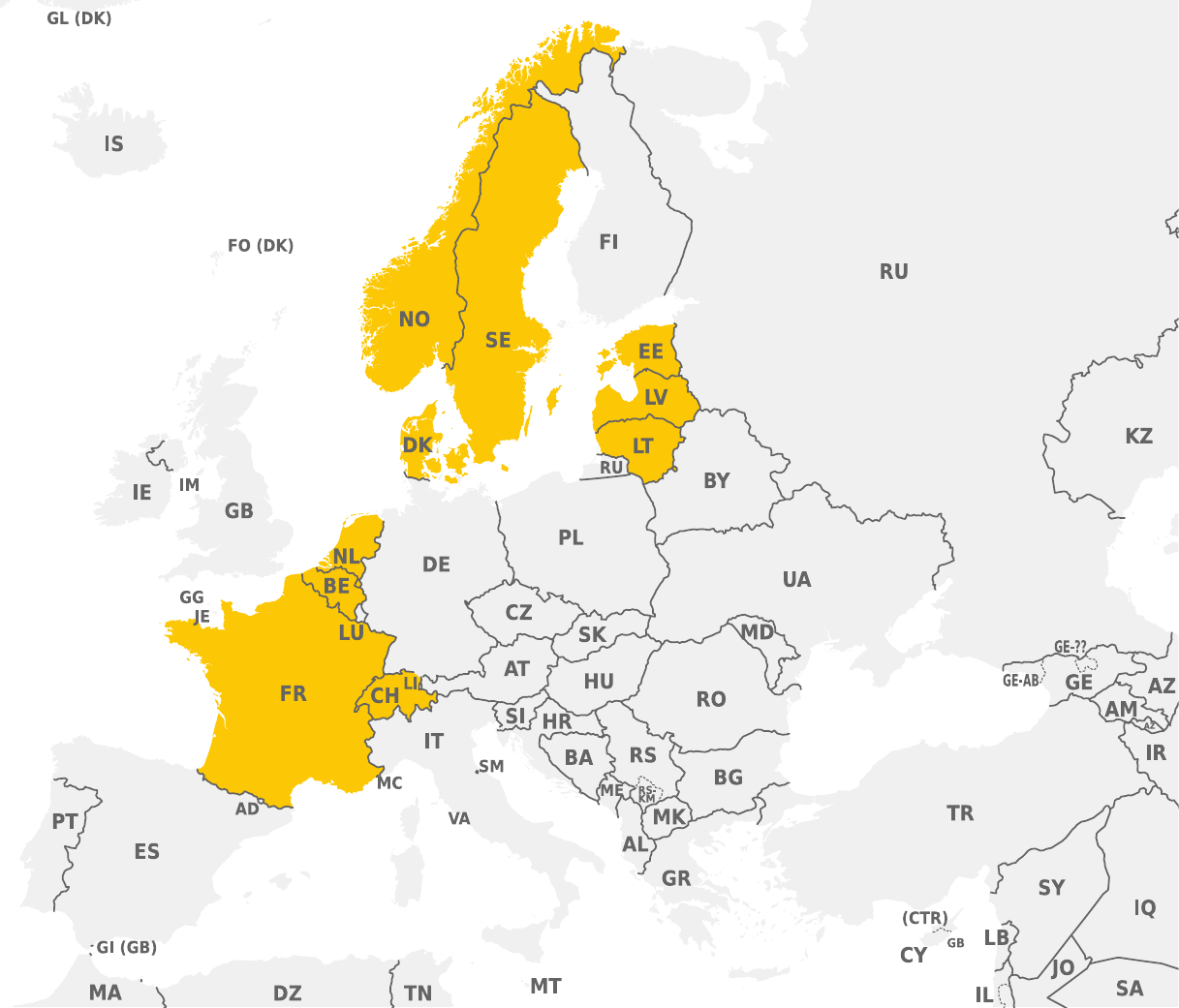}
	\centering
	\caption{The countries we have looked at highlighted in yellow.}
	\label{fig:map-europe}
	%TODO Fix numbers
\end{figure}

% Betere overgangszinnen
In this section we analyse and quantify the impact of the vulnerability we described in the previous section in practice. We limit ourselves to governmental and critical infrastructure domains in the countries as listed in Figure~\ref{fig:map-europe}. The definition of critical infrastructure differs from country to country, but generally encompasses infrastructure, transport, communications, financial institutions, universities, hospitals, etc. We chose these countries based on their score according to ENISA~\cite{enisa_2023} as well as data availability.

\begin{figure}[]
	\includegraphics[width=0.9\linewidth]{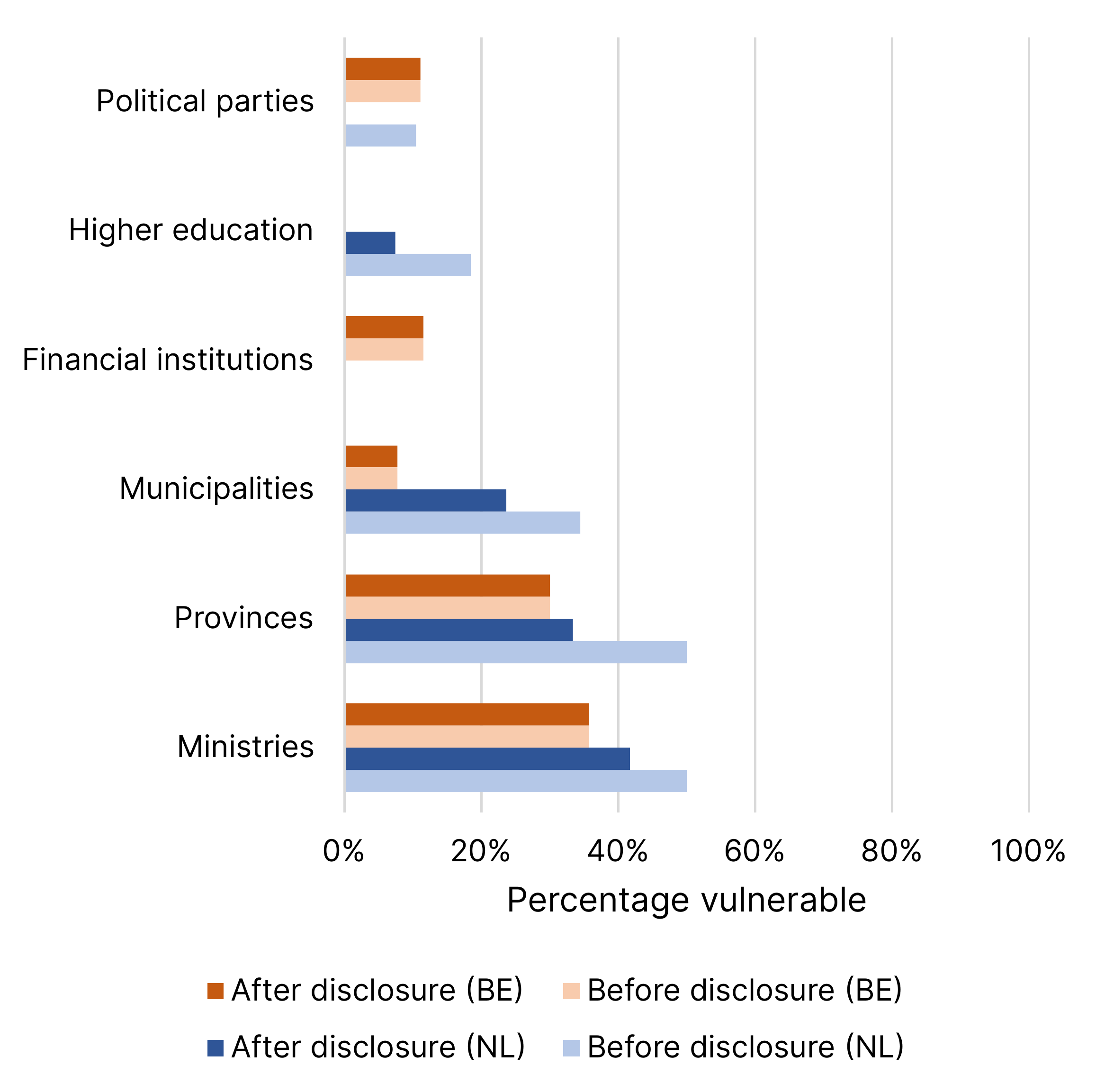}
	\centering
	\caption{The percentage of vulnerable organisations, compared between before and after the disclosure. There were at least 60~days between disclosing and checking again.}
	\label{fig:organisations}
	%TODO Fix numbers
\end{figure}

We tried to validate our heuristic with seven web hosting providers. We selected these providers based on the fact that they are used by government bodies and critical infrastructure. We validate whether our assumptions hold by testing against two classes of web hosting providers. The first type explicitly states on their website that they enforce strict sending policies on their mail servers. We use Dutch hosting provider TransIP as our test subject in this category. The second type does not state anything about sender verification on their website, we select six others in this category. As these web hosters are still vulnerable to these attacks, we have omitted their names. We were able to create a successful proof of concept for all of the hosters in the second category. We tried creating a proof of concept for TransIP, but were unable to do so. As during our search we noticed that most web hosters use similar setups, we refrain from additional experiments with other hosters as these are unlikely to add much value. 

\begin{figure}
	\includegraphics[width=0.9\linewidth]{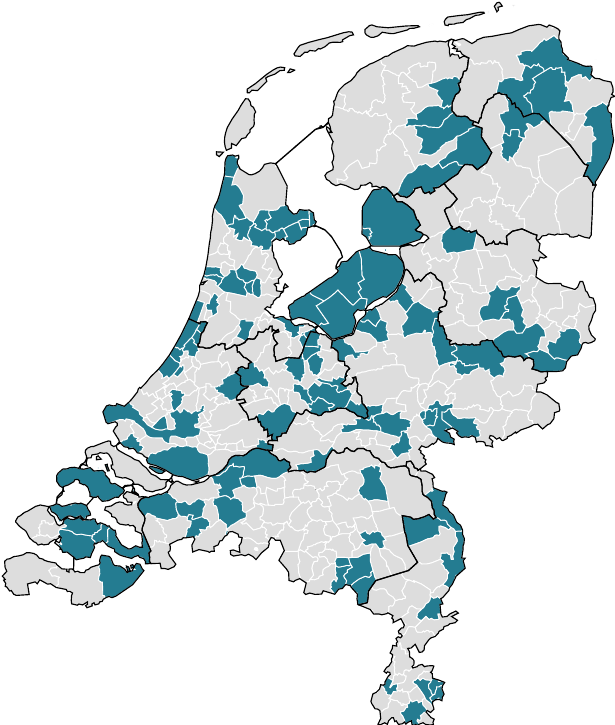}
	\centering
	\caption{A map of all the municipalities of the Netherlands -- the municipalities that are coloured in are the ones we have attempted to contact regarding the vulnerability disclosure.}
	\label{fig:municipalities}
\end{figure}

\subsection{Belgium}
We identified at least 45 likely vulnerable municipalities (out of 581), as well as 3 provinces out of 10. The total number is likely higher. We also identified 3 vulnerable financial institutions. At the federal government level, we found at least 8 well-known federal public services to be vulnerable. On a federal level we ran into problems as there is no publicly available list of federal government domains.
We requested such a list from the Chancellery of the Prime Minister, but received no response. Nevertheless, we can conclude that the issue is widespread at every level of government. In terms of hosters used, we observe a high usage of local national providers along with large international players such as Microsoft. Vulnerable organisations invariably relied on one of the local hosters. 

\subsection{The Netherlands}
Out of the 12 ministries in the Netherlands, we could send email on behalf of 6. Out of the 12 provinces in the Netherlands, we could send email on behalf of 6. Out of the 21 regional water authorities (Waterschappen), we were able to spoof 6. We also looked at all 342 municipalities in the Netherlands. Around ~30\% (114) of them were vulnerable whilst having set a proper SPF (and in most cases DMARC) record. Just like in Belgium, the issue is very prevalent at all levels of government in The Netherlands as well. Their hosters consist of a mix of self-hosted servers, providers specifically aimed at the public sector, and generic web hosting providers, all for the same domain. These organisations tend to be either internationally minded (e.g., Microsoft), or providers from the Netherlands, similar to the situation in Belgium. Many domains share the same providers, especially if the organisations had shared management or worked together in other ways, as is the case for ministries. When looking at the vulnerable ministries, we see that they tend to authorise approximately 25 third parties, out of which we managed to confirm one as vulnerable.

On a central government level, we found a significant number of vulnerable organisations. Providing a percentage is difficult due to the opaque nature of central government organisations and their use of domain names (e.g., use of more than one domain, or use of subdomains). We have attempted to obtain the list of domains used by central government to send and receive email using a Freedom Of Information (FOI or WOO) request. Initially we received an informal response stating that this information cannot be shared as a matter of state security. When we asked for a formal reply, we received a letter stating that this information is not available and thus cannot be shared~\cite{rijksoverheid_2023}. However, we were able to discover at least 8 vulnerable domains that are in the critical infrastructure.

The current numbers show that this issue is widespread. This is a serious threat, as it allows anyone with ill-intent to pose as a government agency. This may be in direct email communication to citizens, or in communication to other agencies. It can be used to make politicians and high-ranking officials say things they never said, and also be used to deny ever having sent actual email. With the current number of impacted government organisations, the potential for abuse is substantial.

\subsection{Rest of Europe}
We have furthermore looked at France, Switzerland, Denmark, Sweden, Norway, Estonia, Latvia, and Lithuania. In these countries it is to our knowledge not mandatory to set up SPF and DMARC properly, and that quickly becomes apparent. SPF and DMARC are by no means as common as in the Netherlands, often at least one of the two is missing. In all countries we could identify at least one government agency that was vulnerable, even if SPF and DMARC were configured correctly. 

The claim that it is not mandatory to do SPF and DMARC is backed up by the responses we received from the national CERTs, who, whilst appreciative, did not consider the issue we reported a concern. For this reason we have left these countries out of our further analysis.

\section{Disclosure \& Response Analysis}\label{sec:disclosure}
We now shift our focus back to the main research question, namely, in this section we analyse what happened when we reported the vulnerability to the various parties we identified in the previous section.

\subsection{Web Hosting Providers}
We first reported the issue we identified to the web hosting providers that are susceptible to this issue. One of them started monitoring and alerting customers which were (unknowingly) using this feature, informing them that it would stop working after a certain date. Some time after that, they started to block spoofed outgoing emails. The other six have not solved the issue. Most of them did not reply or claimed it was `by design'. One of the six even required us to sign a non-disclosure agreement, stating that we would be allowed to go public `at most two months after the issue had been resolved'. This has the caveat that if the issue is never resolved, we would never be allowed to go public. We still informed them without agreeing to those terms via their security email address, but received no response. 

We immediately noticed that security issues raised containing either SPF or DMARC are often dismissed. In many cases `misconfiguration of SPF or DMARC' is excluded from their scope, and whilst we argue that this is not a misconfiguration, the initial assessment often considers it as such. For example, from one bug bounty team, i.e. the team that makes the first assessment of the vulnerability, we swiftly received an `out of scope' reply. This did not change even after we explained to them that this affected all customers of their hosted services products as well.

What makes this vulnerability disclosure difficult is that the third-party email providers have the technical means to solve the issue, but no requirement to do so, whilst the organisations that use those third-party providers that are required to implement email securely, for example due to legislation or binding policy, cannot do so without cooperation from their third-party provider.

\subsection{Organisations}
Given the types of organisations where we observe this vulnerability, and the potential impact of abuse in which an attacker could impersonate, say, a person working at a ministry or other government body, we then decided to approach vulnerable parties directly. Additionally, we had reason to believe that these organisations may disagree with the assessment made by their third-party provider that it is not an issue.

For every organisation, we tried to find a security contact address on their website. If we could not find one, we would email the general contact address. If the organisation was covered by a larger incident response organisation (e.g., a national CSIRT), we contacted that organisation instead. 

We received similar SPF/DMARC dismissals from the national CSIRTs, for example from the Dutch national CSIRT, stating (translated): ``We were able to confirm the problem from your report, but do not consider this a security problem. We will therefore not follow up on it.''. For this reason, we contacted organisations covered by these national CSIRTs directly. 
%From those organisations we have received replies thanking us for our email, and those issues have now been fixed. 
This turned out to be much more productive and led to replies thanking us and the issue being resolved.

We noticed a stark difference in the response time between organisations that are required to set up SPF and DMARC securely. It is obligatory in the Netherlands for governments and critical infrastructure \cite{forumstandaardisatie_spf, forumstandaardisatie_dmarc}, but we could not find similar guidelines for the other countries we looked at.

\subsubsection{Belgium}
As Belgium does not mandate a secure SPF and DMARC configuration, we aimed to compare the vulnerability level and solve rate between the two countries. This is shown in figure \ref{fig:organisations}. The response rate shows a similar pattern: out of the 48 emails we sent to Belgian provinces and municipalities, we received a total of 1 reply within six months (excluding auto-replies), but this resulted in no action. Out of the three financial institutions we contacted in Belgium, none have resolved the issue or responded to it by email after three months, although one did ask us to stop contacting them. We have also noticed that for one bank that when we called to report the vulnerability, nobody within the organisation seemed to know the way to the right department, forwarding us from the reception to the international fraud department, to IT, only to end up being asked to email the general contact address from which -- to this day, over half a year later -- we have not received a reply. On a federal government level we have had no reply after the confirmation of receipt of the list we sent, nor do we see any changes. Our direct emails to the affected departments go unanswered, as did our emails to Belgium's national CSIRT.

\subsubsection{The Netherlands}
For Dutch municipalities, the Association of Netherlands Municipalities (Vereniging Nederlandse Gemeenten, VNG) acts as a CSIRT for all municipalities, which is called the Information Security Service of Dutch Municipalities (Informatiebeveiligingsdienst, IBD). We contacted the IBD by phone and email, and whilst they told us they would contact the affected municipalities, no email to any municipality was sent after 3\textonehalf~months, after which we contacted the municipalities ourselves.

Out of the initial 37 emails we sent to Dutch municipalities, we received a reply for 16 of them. The number of actual replies is lower, as many municipalities work together when it comes to IT (e.g. the municipalities of Haarlem and Zandvoort), thus we received only one reply on behalf of multiple municipalities. The types of responses we received were twofold: one was ``thank you, we will resolve it'', whereas the other one was ``thank you, but we use those services from the third-party and thus we cannot do anything about it''. Only in one case was a system administrator able to convince a web hoster to change their configuration to disallow sending email on behalf of other customers. Another party we contacted in turn contacted their web hoster, who claimed it was not possible. We contacted that web hoster as well, who -- predictably -- said it was by design. 

\subsection{The Case of TOPdesk}
During the last disclosure in the Netherlands, we received a message from the IT manager of one of the affected municipalities, who tested it with its other suppliers as well, and found that a large enterprise service management software supplier called TOPdesk had the same vulnerability in their software-as-a-service solution. This supplier is used by a lot of government entities on all different levels. We initially called TOPdesks's support desk, after which we were recommended to send an email. The email we sent created a ticket, which was promptly closed and marked as `working as intended' the morning after. After pushing further, it would be looked into, but it was still considered user responsibility. TOPdesk maintains an optional blocklist, where an organisation can request that their domain is only usable for their tenant. Based on information we received from a partner, we found a way to scan whether the domain was on the blocklist based on the existence of a CNAME on the organisation's domain. We omit the specifics in this paper to prevent abuse. Our results show only ~20\% of organisations have this set. 

\begin{figure}
	\includegraphics[width=0.9\linewidth]{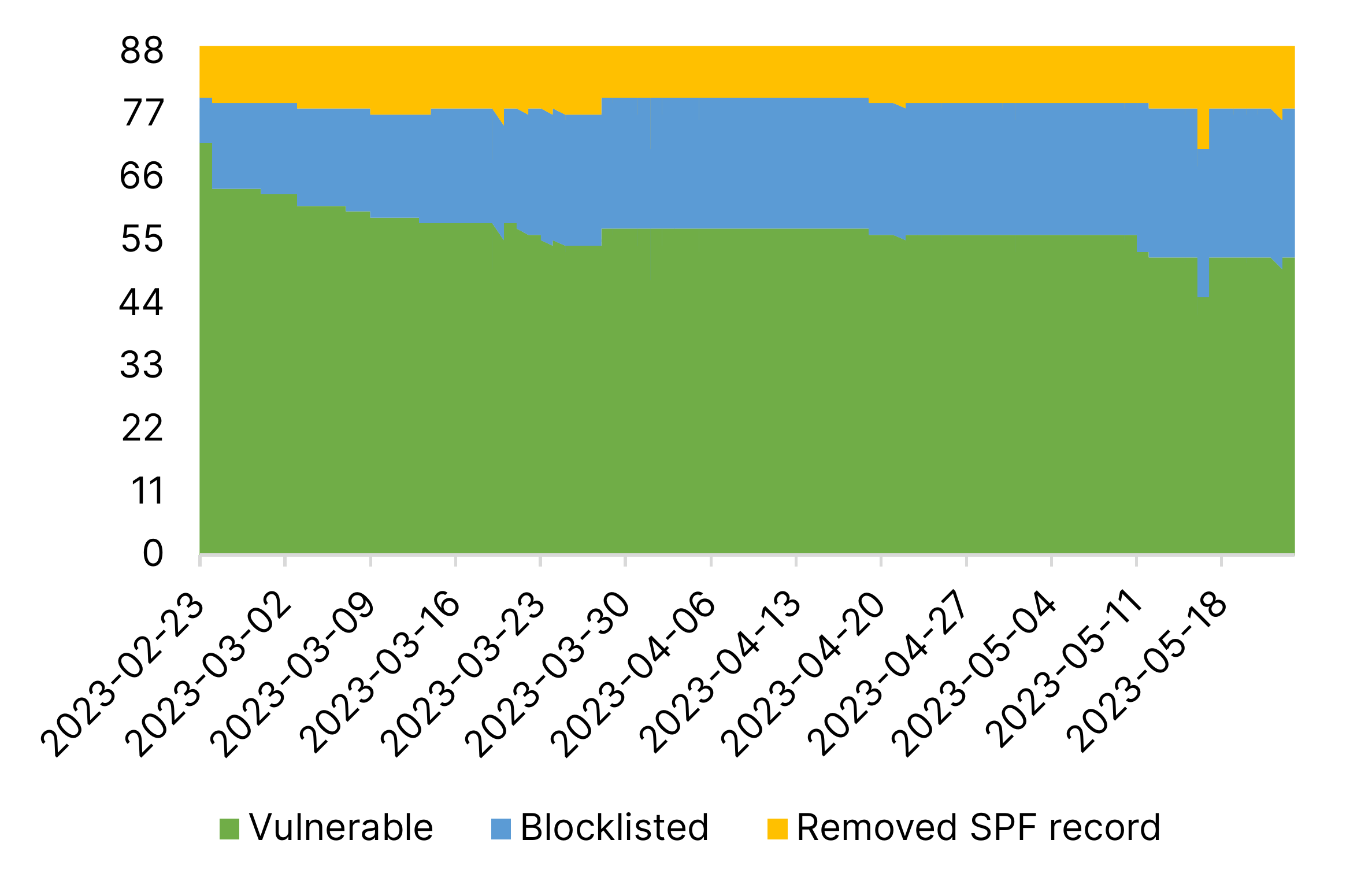}
	\centering
	\caption{Graph for the response time for the second round of disclosures at Dutch municipalities (total 89). Green means they are vulnerable, blue that they requested their domain to be blocklisted by the supplier, and yellow that they removed the supplier from their SPF record. We graphed the time of at least 90 days.}
	\label{fig:topdesk}
\end{figure}

We contacted the municipal CSIRT regarding this. TOPdesk later sent an email to at least some, but not all, of the municipalities. We contacted the 89 affected municipalities on the 22\textsuperscript{nd} of February, 2023, and graph their response times in Figure~\ref{fig:topdesk}. The template of the emails we sent to those municipalities read the following (translated from the original Dutch version):
\begin{quote}
	Dear municipality of \$name,\\
	\\
	From my analysis, it appears that @\$domain has authorised TOPdesk SaaS to send email, without setting any domain restriction. By default, TOPdesk does not check which domain name is used as sender for email sent via TOPdesk. This means that any TOPdesk customer can send email on behalf of @\$domain by simply modifying the "sender" field. This email is indistinguishable to the naked eye from email actually sent by the municipality of \$name, and will therefore pass the technical checks (SPF/DKIM/DMARC).\\
	\\
	TOPdesk is aware that this is possible (ticket X0000-000), but has indicated to me not to change the default settings but to inform the parties. Since it therefore remains possible without action by the municipality of \$name, I am contacting you directly. The IBD has already been informed.\\
	\\
	It can be resolved by having email from TOPdesk sent via your own SMTP server and removing TOPdesk from your SPF record, or by requesting domain restriction from TOPdesk. For this, see also KI 11992 in TOPdesk. This involves any domain name that has include:spf.topdesk.net in the SPF record or otherwise has the IP addresses of the TOPdesk SaaS in the SPF record. I am guessing you have a better overview of this than I do, but in any case it concerns @\$domain.\\
	\\
	If you need more information, please contact us.\\
	\\
	Cordially,\\
	\$reporter\\
	E: \$email\\
	T: \$phone\\
	A: \$postaddress
\end{quote}

\begin{figure}
	\includegraphics[width=0.9\linewidth]{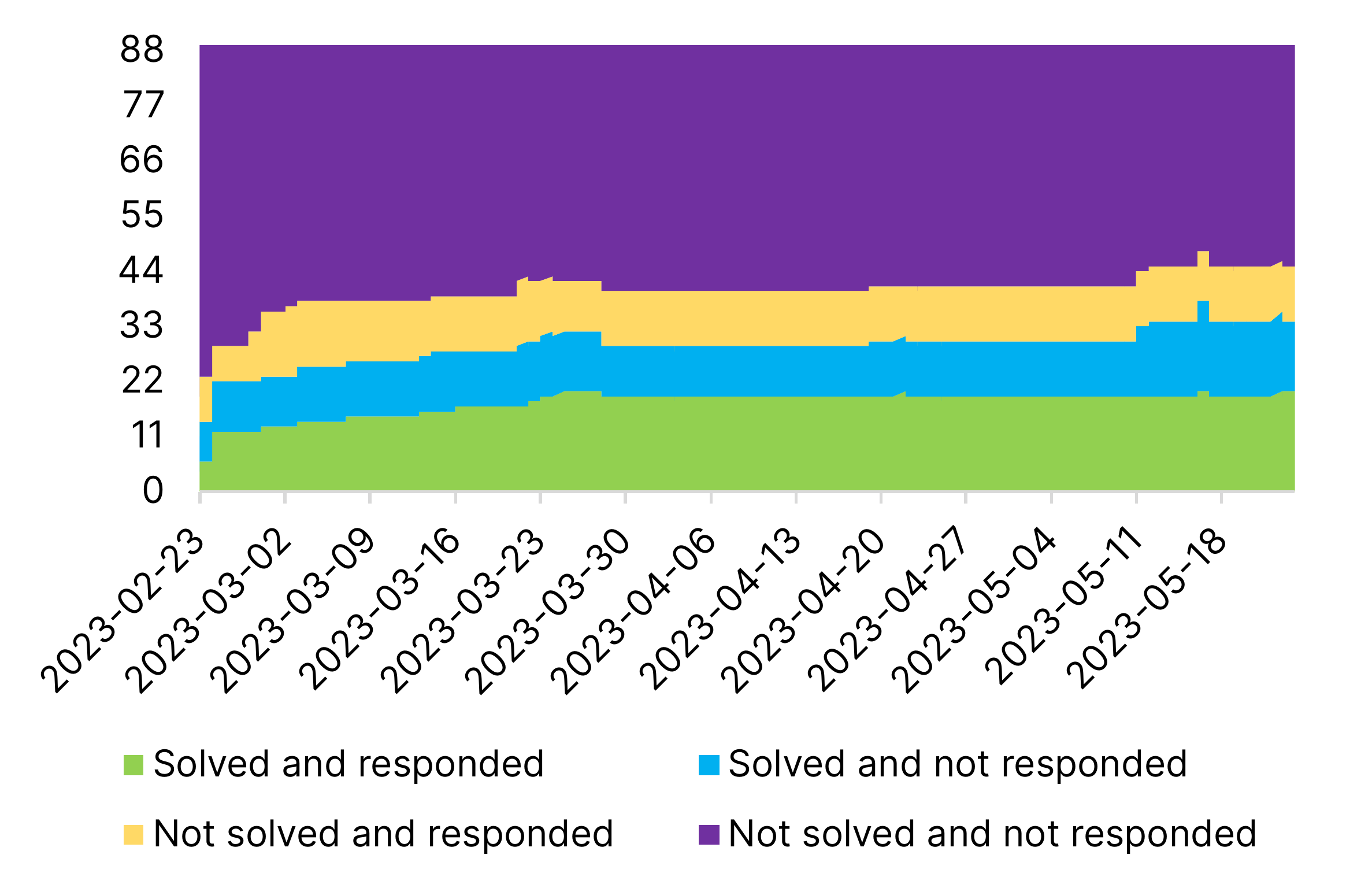}
	\centering
	\caption{Graph for the reply time for the second round of disclosures at Dutch municipalities (total 89). Green means they solved and replied to our report, blue that they solved it but did not reply to us, yellow means that they replied to us but did not solve it, and dark green that they neither replied nor solved the issue.} 
	\label{fig:municipalities-responses}
\end{figure}

For this second round of disclosures at Dutch municipalities we also received responses via email. We can subdivide the municipalities into four quadrants: 1) solved the vulnerability and responded to our email; 2) solved the vulnerability but did not respond to our email; 3) did not solve the vulnerability but did respond to our email; 4) did not solve the vulnerability and did not respond to our email. We graph their responses over time in Figure~\ref{fig:municipalities-responses}. We excluded automated replies from ticketing systems and out-of-office notifications from the responses category. We see that surprisingly there does not seem to be a strong correlation between replying to our message and resolving the issue.

Strangely enough several municipalities removed the supplier from their SPF record, but later added it again. We are unsure why, but we have reason to believe that they intended to use their own SMTP infrastructure with TOPdesk, but after running into issues reverted to using the TOPdesk infrastructure. We base this on the information in Figure~\ref{fig:topdesk}.

\begin{figure}
	\includegraphics[width=1.0\linewidth]{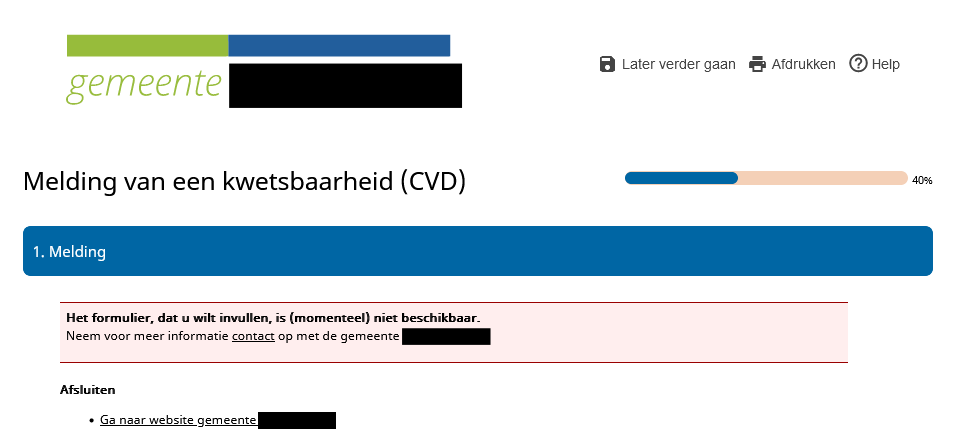}
	\centering
	\caption{A municipality links to a form to report data leaks and security incidents. The form however is no longer available.}
	\label{fig:form-unavailable}
\end{figure}

\begin{figure}
	\includegraphics[width=1.0\linewidth]{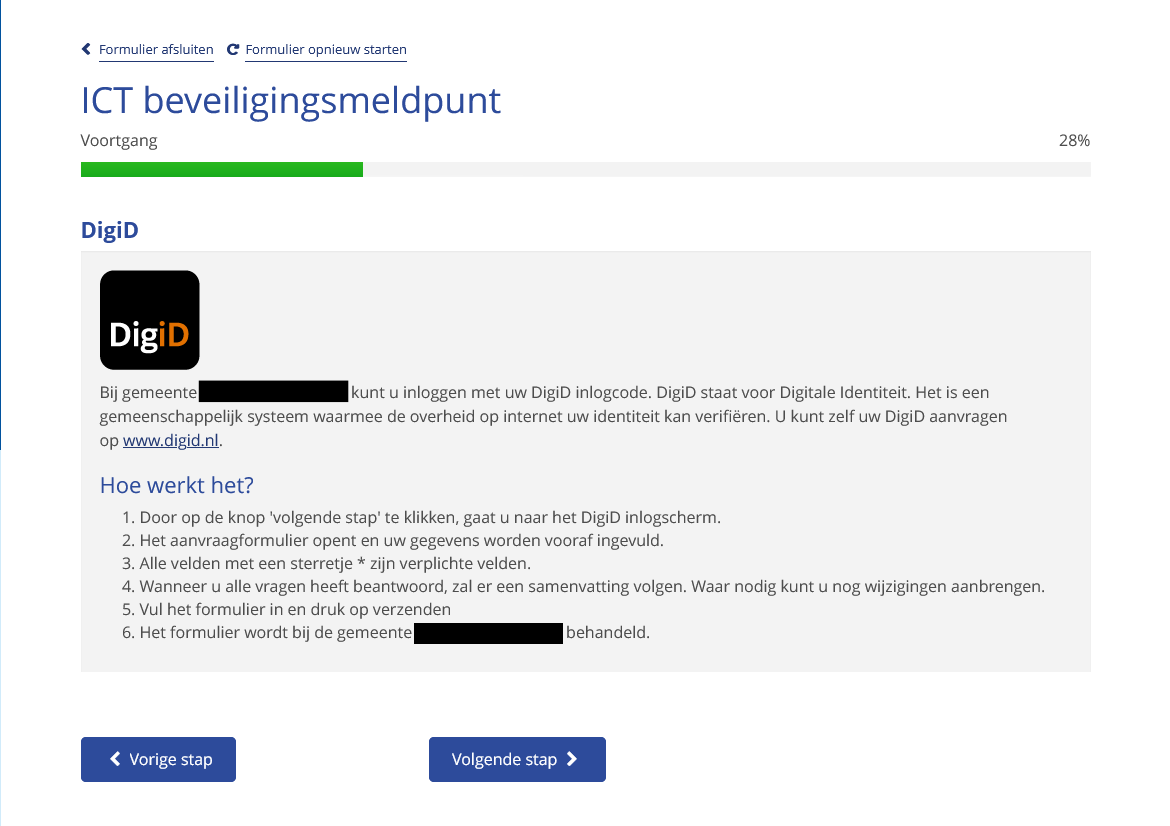}
	\centering
	\caption{A municipality requires signing in with DigiD, a Dutch government system to identify its citizens and inhabitants online, to report a data leak or security vulnerability.}
	\label{fig:form-digid}
\end{figure}

\begin{figure}
	\includegraphics[width=0.9\linewidth]{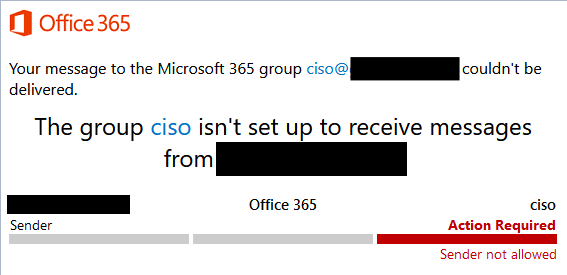}
	\centering
	\caption{The municipality has a clear vulnerability reporting page that asked to email the details to ciso@. Unfortunately the email address listed on that page was not contactable outside the organisation.}
	\label{fig:email-unreachable}
\end{figure}

\section{Policy versus Practice}\label{sec:policy}
We now turn our attention to policy on dealing with vulnerability disclosures. Specifically, we compare the policy of municipalities on paper to what they do in practice (as reported in the previous section).
%We will now examine the policy of municipalities, and compare this to what they did in practice, as can be seen in the previous section, specifically whether their actions match what they say they do. 

In the Netherlands all levels of government (national and local) must implement a coordinated vulnerability disclosure procedure~\cite{bio_2020}. We sent out 114~Freedom of Information (Wet Open Overheid, WOO) requests to the municipalities we attempted to contact for this coordinated vulnerability disclosure and asked them for their policy (if they had one). The exact request is the following (translated from the original Dutch text): 
\begin{enumerate}
	\item The policy documents in force regarding the handling of third-party notifications of security breaches (also known as the `responsible disclosure' or `coordinated vulnerability disclosure' policy), including the policy documents regarding the data stored on the notification and notifier in a `coordinated vulnerability disclosure' notification, as well as all previous iterations, including draft versions, of these policy documents as of the policy document in force on 1~July~2022;
	\item All communications, memoranda, resolutions, information and decisions, and other documents relating to the creation of the current policy or earlier versions of this policy, including all considerations not ultimately included in policy documents, from the policy document in force on 1~July~2022, as well as future versions of this policy;
	\item All reviews, both internal and external, on all iterations of the policy, including draft versions of policy documents that were never published, as of the policy in force on 1~July~2022, past, present, and future;
	\item In the absence of current policies, all communications, memoranda, resolutions, information and decisions on the possible establishment of future policies.
\end{enumerate}
We sent these requests by email, unless the municipality explicitly stated that they do not accept requests via email, in which case we sent a letter via the postal service. 

In all cases the contents were the same. The municipalities we sent these requests to can be seen in Figure~\ref{fig:municipalities}. We received a reply from all contacted municipalities within 10 weeks.

\begin{figure}
	\includegraphics[width=1.0\linewidth]{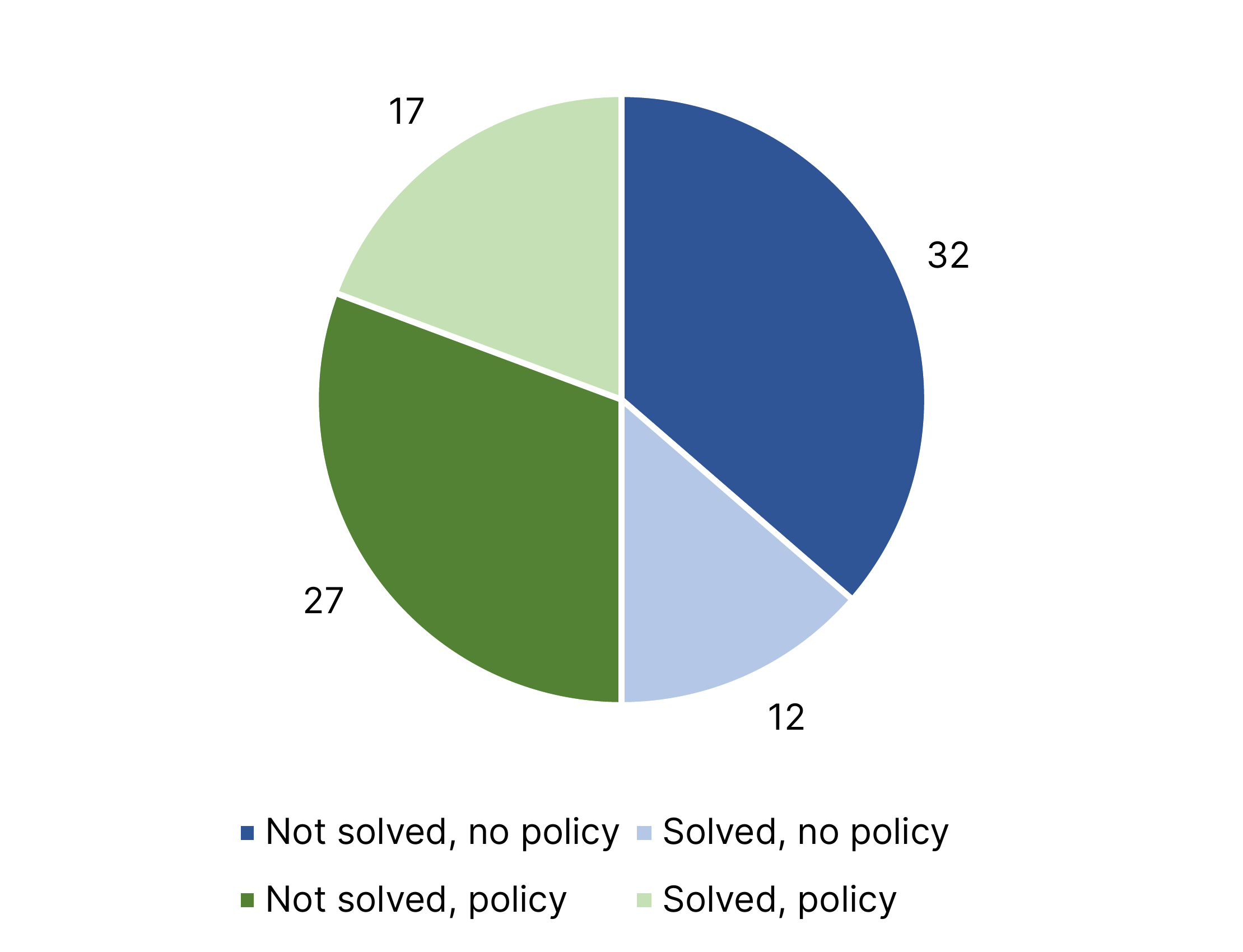}
	\centering
	%\caption{Whether a municipality had policy regarding coordinated vulnerability disclosure notifications compared to their solve rate using the same municipalities as figure \ref{fig:topdesk} and  \ref{fig:municipalities-responses}.}
	\caption{Plot showing whether a municipality had a policy regarding coordinated vulnerability disclosure notifications versus whether or not they solved the issue we reported for the same set of municipalities as in Figures~\ref{fig:topdesk} and~\ref{fig:municipalities-responses}.}
	\label{fig:policy}
\end{figure}

For the TOPdesk vulnerability, based on the data from the WOO requests, exactly half of the 88\footnote{One municipality has ceased to exist since the initial disclosure, and has thus been excluded.} had a coordinated vulnerability disclosure policy. For the group that did not have a coordinated vulnerability disclosure policy, over 70\% of the reports went unsolved. Unfortunately, the situation is only slightly better for the group that did have a policy, with a solve rate of just under 40\%. Our findings are shown in Figure~\ref{fig:policy}. We believe that one of the causes for relatively disappointing solve rate in the group with a policy is the lack of communication regarding the policy. Only in 34 out of the 88 replies to our WOO request did we receive internal communication regarding coordinated vulnerability disclosure. Any unwritten communication is not covered by the WOO, hence we cannot make any statements regarding that. In several cases where the municipality did not have a coordinated vulnerability disclosure policy (CVD), nor any discussion regarding the forming of a policy yet, the WOO request triggered the creation of a CVD policy.

In addition to the above, we noticed that quite a few municipalities added coordinated vulnerability disclosure pages after our disclosure, although sometimes behind a login (e.g., Figure~\ref{fig:form-digid}). 

\subsection{Personal Data} \label{sec:personal-data}
\begin{figure*}
	\includegraphics[width=0.9\linewidth]{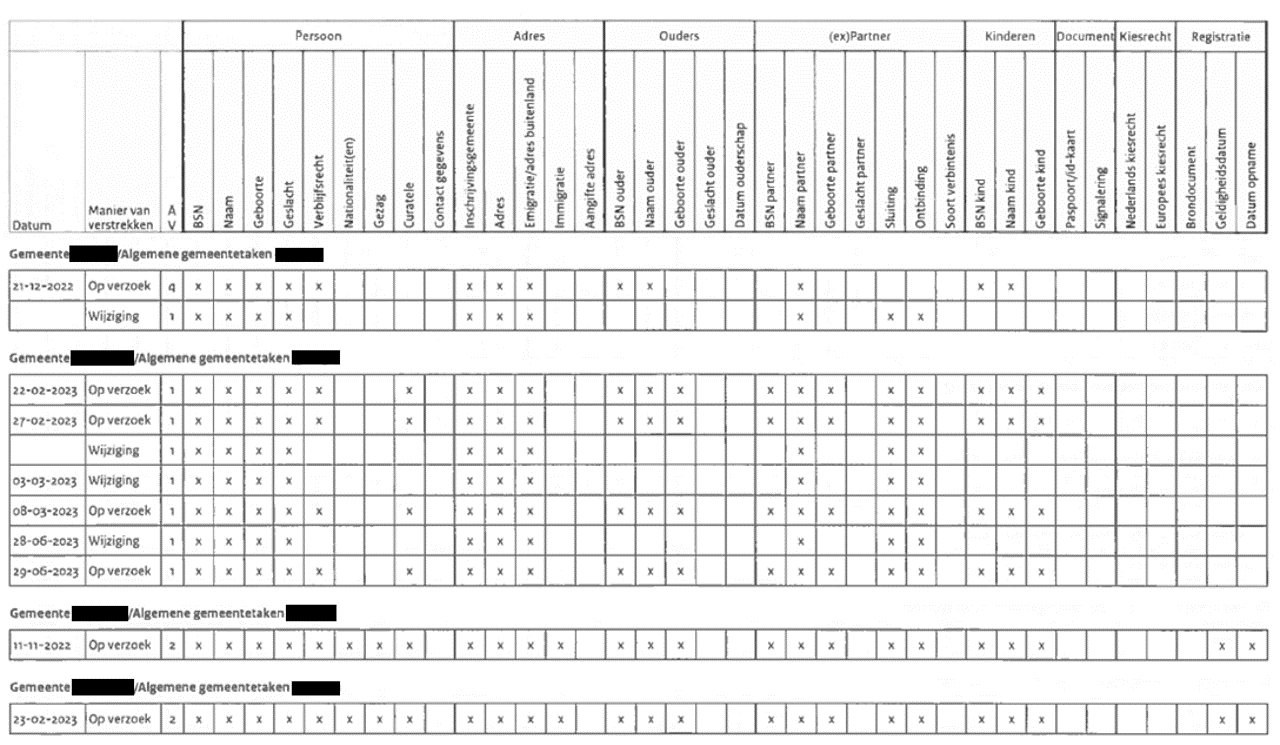}
	\centering
	\caption{A redacted excerpt from the `Overzicht gegevensverstrekkingen uit de basisregistratie personen' (Overview of data provisionings from the Personal Records Database) of the reporter of the vulnerabilities showing four municipalities (`Gemeente') along with what data they requested.}
	\label{fig:brp}
\end{figure*}

During the disclosures, we noticed that in several cases the receiving municipality would look up the personalia of the reporter. We saw, for example, lookups via the social media platform LinkedIn. Government organisations in the Netherlands can request data about citizens from the Personal Records Database (BRP). Every one of those requests is logged by the National Office for Identity Data (RvIG), and an overview of the requests made for one's data can be requisitioned by every citizen. As we suspected that some municipalities might be using that data after we performed our disclosures, we requested such an overview. Around the date of the disclosure, 11 municipalities requested data from the BRP, as can also be seen in Figure~\ref{fig:brp}, including in one case: 
\begin{enumerate*}
	\item citizen number;
	\item name;
	\item date of birth;
	\item gender;
	\item residential status;
	\item nationality;
	\item authority;
	\item guardianship status;
	\item registered municipality;
	\item address;
	\item emigration status and address;
	\item migration status;
	\item citizen number of the parents;
	\item name of the parents;
	\item date of birth of the parents;
	\item citizen number of the registered partner;
	\item name of the registered partner;
	\item date of birth of the registered partner;
	\item start date of the registered partnership;
	\item end date of the registered partnership;
	\item citizen number of the children;
	\item name of the children;
	\item date of birth of the children.
\end{enumerate*}

Requesting data from the BRP registry requires a legal basis for processing data (per the GDPR \cite{eu-679-2016}). As far as we now, and given the content of the policies we obtained, no such basis exists in case of a coordinated vulnerability disclosure.

We contacted the eleven municipalities that requested data from the BRP directly and asked them for the reason they requested that data. To our surprise, most municipalities were unaware that they requested that data. We received one reply from a municipality stating that they always request this data for any form, and then throw out the information that they do not need, which in this case was everything about the parents, partner, and children. This data is then, according to their policy, stored for five years after the notification. In another case it was a tick mark that was accidentally turned on, thereby requesting far more information than was necessary. We have also received two replies stating that the grounds on which the data was processed was `illegitimate'. The other replies can be summarised as ``the system required the information'', with data generally being stored for one to ten years.

As the citizen number, name, and date of birth of the reporter's parents were also requested, the reporter asked their parents to also request the information which organisations retrieved from the BRP. In nine cases, their data was requested as well, including:
\begin{enumerate*}
	\item citizen number;
	\item name;
	\item date of birth;
	\item gender;
	\item residential status;
	\item registered municipality;
	\item address;
	\item emigration status and address;
	\item migration status;
	\item citizen number of their parents;
	\item name of their parents;
	\item date of birth of their parents;
	\item citizen number of their registered partner;
	\item name of their registered partner;
	\item date of birth of their registered partner;
	\item start date of the registered partnership;
	\item end date of the registered partnership;
	\item citizen number of their children (which includes the reporter);
	\item name of the children;
	\item date of birth of the children.
\end{enumerate*}
As the reporter's grandparents are no longer capable of requesting their data, we are unable to tell whether their data was requested as well. When we asked why this data was requested at the municipalities based on our rights laid out in the General Data Protection Regulation (GDPR), we received at least one response that this data had not been requested even though we can see that it has been requested. After we contacted the RvIG, who in turn contacted that municipality, this was changed to `the data is not being used'.

It is unclear which data a municipality is legally allowed to request on the reporter of a coordinated vulnerability disclosure notification. As circa 90\% of municipalities show, no additional information is necessary in principle. However, there is to our knowledge no jurisprudence regarding this matter. The IBD states that excess data requests are not permitted~\cite{ibd_2023}.

\section{Discussion}\label{sec:discussion}
In this section we elaborate on our experiences as described in the previous sections; we have seen that disclosures do not always go smoothly. 

\subsection{`Email is Insecure'}
Whilst email is the most prevalent digital communication medium, used for messages that require security and a guarantee of authenticity, it seems that the prevalent notion of `email is insecure' prevents adoption of technologies to make email more secure. Especially given that in the case of the vulnerability we reported there is a chain dependency on third-party suppliers, where a single insecure link in the chain makes the entire chain insecure, this notion seems exceedingly harmful.

Additionally, we note that web hosting providers tend to initially answer either that the vulnerability we report does not exist in their systems, or that it is by design, leaves organisations that do want to resolve the issue with only two options: 1) convince the webhosting provider to resolve their issues; 2) stop using the web hosting provider. Since these decisions are often made on a long term basis, the latter option is often infeasible. Out of all the web hosters we tested and verified, only one has changed their configuration, showing a lack of urgency on their side.

We also found that for a lot of organisations outside the Netherlands, SPF and DMARC were not deemed important, where the configured policies were lax with a \verb|+all|, or in the case of the United Nations (un.int) at one point even explicitly granting \verb|0.0.0.0/0| (i.e. the entire IPv4 space) permission to email on behalf of them. We did not contact these organisations unless they were obliged by law or treaty to implement SPF and DMARC properly.

On top of this we were surprised to find that organisations that would benefit the most from a secure SPF and DMARC policy (e.g., financial institutions who advertise in bold letters on their website to look out for phishing) were the most hesitant to resolve the issue.

All-in-all it seems like the adoption and proper application of SPF and DMARC is very much dependent on either regulation or the enthusiasm of a couple individuals within an organisation. 

\subsection{Reporting Difficulties}
We noticed a stark difference between the Netherlands and other countries in terms of the availability of a coordinated vulnerability disclosure contact address, as well as response to our notifications. We believe this to be due to the policy push by NCSC-NL in 2013 \cite{kinis_2018}, although we do not rule out the influence of the nationality of the reporter matching the nationality of the organisation that is reported to. Still, we encountered several cases where the security contact address no longer existed, the displayed address was different than the actual linked address, or the email was reported as not received by a follow-up phone call, or we only got in contact with the security team because a colleague knew someone, or we managed to guess the email address of the CISO based on the common email pattern for the organisation and a LinkedIn profile for said CISO. The teams responsible for security within organisations are often unaware that they are unreachable, as the message does not arrive, and first line support is often unprepared to handle these cases. 

During the disclosure to municipalities, we found that it could be difficult to reach the municipality, even if they had a coordinated vulnerability disclosure page. For example, some had a coordinated vulnerability disclosure submission form that was no longer available (e.g., Figure~\ref{fig:form-unavailable}), some required signing in with a government ID, barring all foreigners from reporting an issue (e.g., Figure \ref{fig:form-digid}), whereas some had an email address that was unreachable from outside the organisation. After our disclosure, we noticed that several municipalities started implementing coordinated vulnerability disclosure pages. At least two have since added a coordinated vulnerability disclosure page that is only accessible behind a sign-in with a government ID similar to Figure~\ref{fig:form-digid}.

Out of the organisations that had a coordinated vulnerability disclosure guideline with response deadline, over half of the times this deadline was not met, for example because the contact address listed on the coordinated vulnerability disclosure page was not aware that they were the contact person, the page listed multiple, generic contact addresses that were not aware of the coordinated vulnerability disclosure process, or the email address listed on the page was ambiguous, where the link of the email and the email displayed on the page differed.

It seems like the reputation of the reporter plays a role for some organisations. We sent the reports as a private individual, but we received some internal messaging that suggests the 
university
%TODO ANON
%REDACTED FOR DOUBLE-BLIND REVIEW
affiliation of the reporter that came up when they searched for the reporter increased the trust in the report.

\subsection{Reactive versus Proactive}
We noticed an uptake in activity once we sent out our freedom-of-information (WOO) requests. We received a response that a municipality restarted the decision process for a CVD policy, and also received reports that our WOO request prompted another look at our initial CVD report. 

Additionally, the preliminary results of our research, specifically those regarding the solve rate at municipalities, appeared in national media \cite{news_agconnect, news_binnenlandsbestuur, news_ibestuur, news_nos}, which also spurred municipalities and other organisations to take another look. We have received several confirmations since the media release that organisations would look into it again.

\section{Recommendations}\label{sec:recommendations}
Given the challenges we encountered in reporting a simple yet potentially serious email vulnerability, we make a number of recommendations to public organisations on dealing with coordinated vulnerability disclosures.

\subsection{Regulation}
As can be seen in Figure~\ref{fig:organisations}, there is a stark difference between the Netherlands and Belgium when it comes to responding to and solving our disclosures. We have noticed patterns similar to those of Belgium in other countries. We want to highlight once more that we only contacted organisations that attempted to properly set up SPF and DMARC, meaning we already have a positive selection bias. We therefore believe that awareness and policy, as is currently implemented in the Netherlands, helps. Even the municipalities that did not have a coordinated vulnerability disclosure policy in many cases did respond to our message. In these cases they were aware of what a CVD policy was, but did not yet come around to creating one. Nevertheless, this led to our message arriving at the right person within the organisation. We believe that requiring this by law or policy, as is the case in the Netherlands as part of the BIO \cite{bio_2020}, can create this awareness. We believe that mandating \verb|security.txt|, as currently mandated by Forum Standaardisatie \cite{forumstandaardisatie_securitytxt} and recommended by NCSC-UK \cite{ncscuk_2020}, can also help in this endeavour.

One thing that stood out for the 44 municipalities that did have a policy is that almost all of them have a policy based on a template supplied by the overarching CSIRT for municipalities, IBD. This creates a solid base for policy within the municipalities, but may potentially also lead to the requirement becoming a `box-ticking' exercise, where insufficient thought is put into the processes behind the policy. We strongly believe that expanding the number of organisations required to implement a coordinated vulnerability disclosure procedure, either by making adherence to ISO 29147:2018 \cite{iso29147:2018} or the EUCC \cite{enisa_2021} a requirement, would greatly improve the notification handling at other organisations. We believe that there is a role for ENISA to increase awareness across the EU. % in other countries in the EU outside the Netherlands.

\subsection{Enforcement}
We have seen that even though attempts have been made to encourage providers to implement solutions that use open standards and make secure solutions, such as the Leveranciersmanifest (Vendor Manifest)~\cite{forumstandaardisatie_leveranciersmanifest} and Veilig Email Coalitie (Secure Email Coalition)~\cite{veiligemailcoalitie}, that in most cases there is no enforcement mechanism to ensure that those who sign the manifest actually adhere to it. This leads to situations where suppliers that an organisation thinks do things correctly or aim to do so, by virtue of having signed on to the document, in reality do not. We urge the organisers of these coalitions to set clear guidelines and processes for the dismissal of participants that do not or no longer adhere to the guidelines set during onboarding.

\subsection{Coordinators}
During our disclosures, we ran into several problems with so-called coordinators -- organisations or teams (e.g., CSIRTs) that relay information from the outside world (i.e., us) to the vulnerable organisations. This resulted in implicit gatekeeping -- in several cases we found that such an organisation (such as the IBD or NCSC-NL) did not relay the information to the vulnerable parties, even though later it turned out the vulnerable party had wanted to act on it. Whilst this filtering may make sense in some cases (e.g., if a vulnerability has very limited impact), it is not part of the coordinator description of ISO/IEC 29147:2018 \cite{iso29147:2018}. We therefore recommend CSIRTs to provide and communicate clear guidelines on what type of reports will be forwarded, and equally important which ones will not be forwarded, both to their members and the ones reporting. We recommend those members to consider carefully whether this filtering is in line with the notifications they wish to receive, and to alternatively provide a means to directly report issues to the organisation.

\subsection{Evaluation}
Additionally, we find that when a coordinated vulnerability disclosure procedure exists, it may not work, both technically and within the organisation. Examples such as Figure~\ref{fig:form-unavailable} are no exception. In many cases reporters may simply give up on reporting vulnerabilities, which leads to a false sense of security. After all, how does one notice that no one sends in a report because the form is broken versus no one sends in a report because there are no vulnerabilities found? We recommend having these procedures tested by an external person or agency. Situations such as shown in Figures~\ref{fig:email-unreachable} and~\ref{fig:form-unavailable} would be picked up during that process. In the answers to our WOO requests we also saw that very few municipalities in the Netherlands currently perform evaluations of their CVD policy. For this reason, we urge both ENISA and the BIO working group to consider making periodic evaluations of the current CVD policy and processes a requirement of the EUCC and BIO respectively.

\subsection{Organisational Awareness}
We recommend checking that those working in the organisation are aware of the department or individuals to forward disclosure notifications to. In some cases it can be difficult to directly find the right contact person, especially if there is no coordinated vulnerability disclosure policy published. We noticed several cases where the recipient was weary of our authenticity, but decided to forward the notification anyway to the security officer. These cases can be seen in Figure~\ref{fig:policy} as `solved, no policy'. Had they not done so, the message would have been lost. We urge organisations to raise awareness of whom to contact in case of security notifications, especially to those with front-facing roles such as the reception, for example by making it part of security training procedures.

\subsection{Anonymity}
When reporting a data leak or software vulnerability, the reporting party may not wish to share all information. We therefore recommend minimising the data requested to report a vulnerability. Gender, date of birth, and citizen number, as well as those details of family members or spouses are generally not required to handle the vulnerability report, even though we saw in Section~\ref{sec:personal-data} that this information is sometimes requested. The municipal CSIRT IBD has stated that requesting and processing this information is not allowed \cite{ibd_2023}. From communication with the parties that did request this information, we have had no case where they claimed to have used this information, whereas in conversations with people reporting CVD issues, we heard that requiring this kind of information plays a significant role in the consideration whether to report an issue. We believe there is a role for the sectoral CSIRTs to highlight the importance of anonymous reporting, and for ENISA and the BIO working group to mandate an anonymous reporting option.

\subsection{Simplicity}
Lastly, we want to highlight that the ones that make the report very often do so without necessarily wanting anything in return. The law in territories such as the Netherlands forbids abusing these vulnerabilities, but does not require reporting them to the organisation or outlaw selling the information to third-parties (so-called grey markets)~\cite{ncsc_2018}. Discoverers could therefore, in theory, opt to sell their vulnerabilities on the `grey market'. We urge organisations to make the process of reporting the vulnerability as simple as possible, and underscore that the organisation where the vulnerability has been found is not in a position to demand anything from the reporter. This includes non-disclosure agreements, providing personal information, adhering to timelines, and more.

\section{Ethical Considerations}\label{sec:ethics}

The study in this paper poses obvious ethical concerns, as it describes vulnerabilities that were and in part still are exploitable. We have taken the utmost care to make sure vulnerabilities were properly disclosed to the parties involved. The security of the organisations and their users was paramount. We believe however that not publishing our findings will in the end do more harm than publishing them. We followed official guidance from NCSC-NL which stipulates that organisations should have at least 60 days to resolve reported issues~\cite{ncsc_2018} -- in practice all organisations we contacted had over 180 days to resolve the issues at hand prior to public disclosure.

Additionally, testing for the vulnerability also came with ethical concerns. To mitigate these, we ensured that we only sent email to addresses we controlled, from email addresses that were clearly identifiable as stemming from us, using message bodies that made it clear that this was a test, unless the organisation it concerned had explicitly given us permission to test other recipients and bodies. We have executed our tests on a limited subset of affected organisations that shared the same configuration based on our scan results in order to minimise harm.

It is possible that our results will be used by malicious parties to send phishing on behalf of the affected organisations. We have therefore tried to reach the organisations and third-parties beforehand, as well as the relevant CSIRTs. We have also presented our findings as a closed (TLP:AMBER \cite{first_tlp}) presentation to the FIRST conference to inform the incident response community first.

Finally, we consulted the computer science ethics board of the 
University of Twente
%TODO ANON
%REDACTED FOR DOUBLE-BLIND REVIEW
, who stated that since the reporting of the vulnerabilities started before the involvement of the university
%TODO ANON
%REDACTED FOR DOUBLE-BLIND REVIEW
, they are unable to comment on that. For publication we have decided to redact as many names as possible. However, due to the nature of this process, redacting all names is not feasible. Throughout the process we adhered to the 
Dutch Institute for Vulnerability Disclosure
%TODO ANON
%REDACTED FOR DOUBLE-BLIND REVIEW
's code of conduct and complied with the 
University of Twente
%TODO ANON
%REDACTED FOR DOUBLE-BLIND REVIEW
's vulnerability disclosure policy.

\section{Conclusion}\label{sec:conclusion}
Throughout this paper, we used an email vulnerability discovered by us to test how public bodies such as municipalities, regional and national authorities respond to vulnerability disclosures. We then compared their handling of vulnerability reports to public policy as set out in the Netherlands. We found that a significant number of government and critical infrastructure organisations are vulnerable to the vulnerability we discovered, even with strict SPF and DMARC records. We have also shown that the disclosure process at several organisations struggles to deal with reports like these. All but one affected web hoster found the issue to not be worth resolving. When we reported the vulnerability directly to affected organisations within government and critical infrastructure, in over half the cases the issue was not resolved, regardless of whether they had a coordinated vulnerability disclosure policy or not. We therefore urge policy makers to take steps in creating effective policy on coordinated vulnerability disclosure.

\subsection*{Data Availability}
All data to produce these results are available on request. The datasets themselves are not published publicly to avoid harm to the affected organisations, as they provide details about the third-parties used by every organisation. The source code is intertwined with these datasets and can thus not be provided separately.

%TODO ANON
\section*{Acknowledgements}
We want to thank the people at the Dutch Institute for Vulnerability Disclosure for their valuable input throughout the process, as well as all the security officers, privacy officers, and others who we have spoken to over the course of this research.

\bibliographystyle{plainurl}
\bibliography{paper}

\end{document}